# Spatiotemporal representation of a two-vortex reconnection as a single rotating vortex

Jordan M. Adams

Riverside Research Institute, Beavercreek, Ohio 45431, United States.

## Abstract

Reconnections and rotations of lines are dual descriptions of the same saddle-shaped spacetime surface. We show that a reconnection between two line occurring over time is a single line that rotates over space progression. Both rotating lines and reconnections possess the same saddle shape sheet geometry in four-dimensional space-time, with different orientations. Cyclic precessing lines occurring over time are arrays of reconnections occurring spatially. We show that a magnetic reconnection occurring over time can be seen as a single continuous line vector potential rotating spatially, where the full evolution traces a saddle shape surface. Finally, we show that a single tilted spatiotemporal optical vortex precesses with spatial progression, and as a result can be seen as two vortices reconnecting. Given the unique spatiotemporal evolution, we also analyzed the relativistic angular momentum of these electromagnetic fields.

## 1 Introduction

Vortex reconnections are events where multiple vortices accelerate toward each other, intersect and trade halves, then accelerate away [1-5]. Turbulence can be driven by reconnections in viscous fluids [1-3] while stellar phenomena like solar flares involve magnetic reconnections in plasma [5]. The distance between two vortices during a reconnection has been found to generally follow a half power scaling law with time. These events involve complicated vortex motion and understanding their trajectories is important for predicting earth and space weather, along with managing acoustics from moving air. It would be valuable to simplify these interactions conceptually as well as better understand the reason behind their scaling motion. Understanding these interactions will unlock physical understanding of a large number of phenomena and may improve predictions of these events.

Recently, reconnections have been demonstrated between spatiotemporal vortices and traditional spatial vortices in electromagnetic fields [6-8]. The vortices are embedded in a wavepacket and reconnect as the field travels. As of yet, there has been no investigation into the angular momentum of electromagnetic reconnections. The transverse angular momentum of spatiotemporal vortices has been strongly debated, where some have suggested an alternative transverse angular momentum operator to ensure conservation, as the standard operator gives an apparent torque, while others have suggested the torque arises only when using paraxial and narrowband approximations [9-13]. Unlike the standard optical vortices, reconnecting vortices evolve over time and it would be interesting to analyze the four-dimensional angular momentum for such fields. For a field in spacetime, the relativistic counterpart for angular momentum density is a tensor with six key components [14,15]. These include the three well-known angular momentum components arising from spatial rotations along with the three additional angular momentum components arising from Lorentz boosts, all of which are conserved quantities for fields in free-space.

In this paper, we show that two vortices reconnecting are equivalently seen as a single rotating spatiotemporal vortex from different four-dimensional perspectives. In Section 2.1 we use a toy model to first present a geometric perspective of null-line evolutions, which represents both the core of vortices or magnetic field lines during reconnections. The null-line plotted over time is represented as a null-surface which is a saddle shape for a reconnection. Seeing two null-lines reconnect versus seeing a single null-line rotating is determined by how the same null-surface is "sliced". Cyclic precession is shown to be an array of reconnections that progress spatially, while the same event in a different inertial frame reveals the precession can also been seen as a reconnection with temporal progression.

Extending from the geometric analysis, we show a two-dimensional magnetic reconnection simulation that corroborates the results in Section 2.2. The magnetic field lines reconnect over time, while only one field line is seen to tilt spatially. The magnetic field line is plotted over time and stitched together into a null-surface, which has a saddle shape.

In Section 2.3, we show tilted electromagnetic spatiotemporal vortices precess with propagation and, as a result, can also be seen as reconnecting electromagnetic vortices. First, we demonstrate that cylindrical focused optical vortices (OVs) and propagating spatiotemporal optical vortices (STOVs) are static reconnections. Next, we show that STOVs with spatiotemporal tilt precess with propagation and are dynamic reconnections due to the vortex rotation. We calculate the angular momentum for the full electromagnetic field vectors and find that analytic fields from paraxial and narrow bandwidth approximations give rise to a fictitious torque, while numerically calculated fields without approximations have angular momentum that is constant with time. Finally, we calculate the intrinsic angular momentum by shifting the calculation to the energy centroids.

We believe this work offers a simplified perspective of reconnections which will aid in predicting and controlling reconnections in air, electromagnetics, and magnetohydrodynamics.

# 2 Results

## 2.1 Geometry of reconnecting lines as a tilting line

While the physics for systems constrains unique relationships between space and time for fields, there are basic four-dimensional geometric properties underlying reconnections across different systems. To present the underlying geometric concept, we start with a toy-model representation of reconnections. The core of the vortex is represented with a line (null line) and magnetic fields are represented with the lines of constant vector potential. Geometrically and regardless of the physics, two lines that reconnect over time are a single line that rotates spatially. A toy reconnection model of a scalar field, which closely resembles that found with light propagation (Section 2.3), is

$$\psi = (a(x^2 - z^2) - vt + iy)u \tag{1}$$

where $a = 1$/mm, $v = 1$ mm/ps, and $u$ includes a slowly varying envelope and fast-varying phase information, which to study the space-time geometry of the null lines, we set $u = 1$/mm. This describes two vortices traveling along $z$, accelerating, reconnecting, then moving apart in $x$ as shown in Figure 1 (a). The null line, $Re(\psi) = 0, Im(\psi) = 0$, can also represent magnetic field lines reconnecting as discussed in Section 2.2. The temporal evolution of the null line is shown in Figure 1 (a). The two lines move along $z$ with the relation $z = t^{\frac{1}{2}}$ at $x = 0$ for $t < 0$, merge into an "X" at $t = 0$, and then move along $x$ with relationship $x = t^{\frac{1}{2}}$ at $z = 0$ for $t > 0$. On the other hand, it can be helpful to visualize the entire null-line evolution as a single space-time surface (Figure 2 (c)). Each time-slice intersects twice, forming two lines.

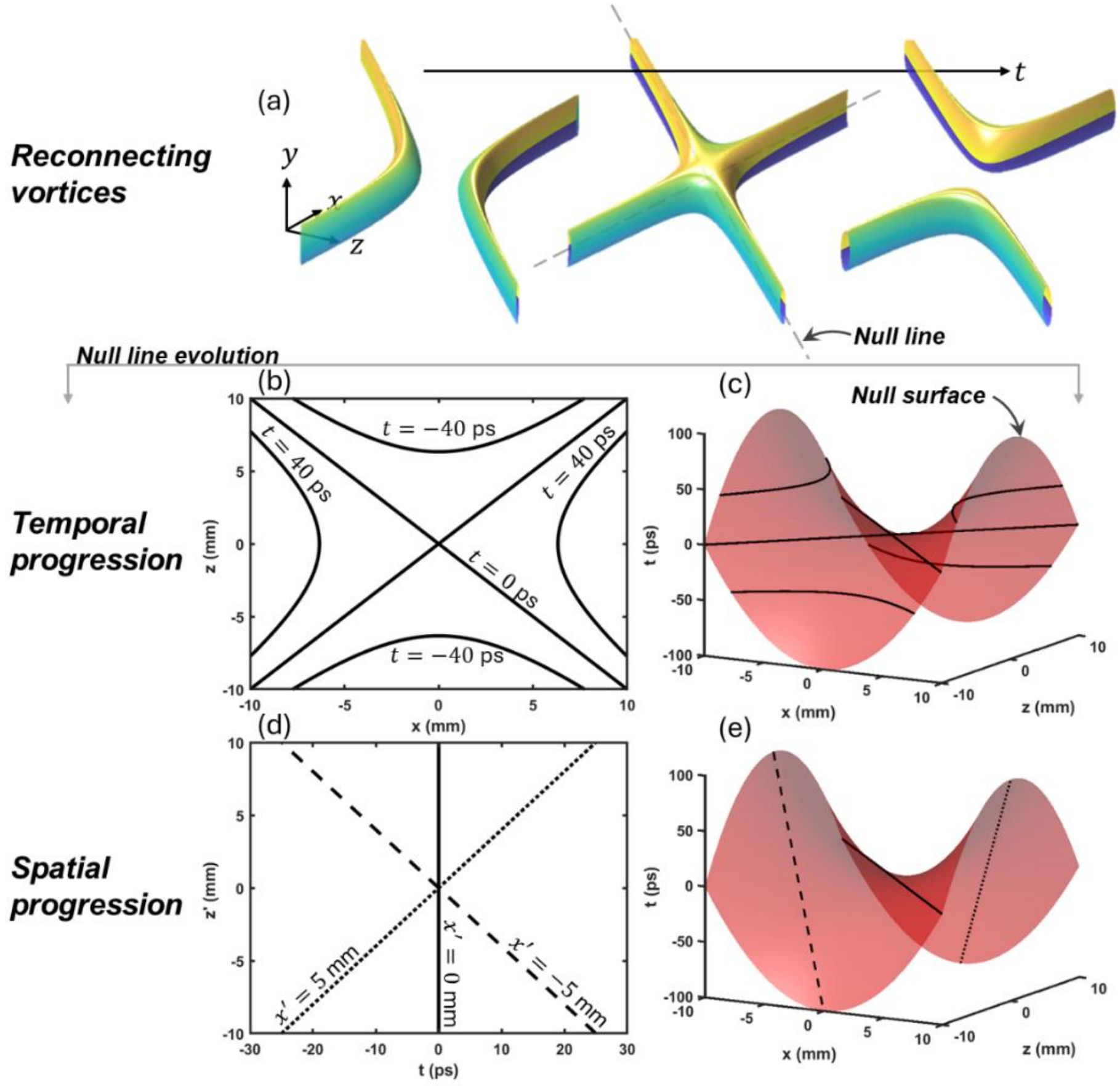

**Figure 1:** Comparing standard temporal progression of reconnecting lines to spatial progression. **(a)** Temporal progression shows two lines bend and accelerate toward each other, form an "X" intersection, then accelerate apart. **(b)** Temporal progression of the null lines or center-line of the vortices. **(c)** The line's full trajectory can be visualized with a surface. Each time-slice intersects this surface and gives two-lines or a connected "X" at $t = 0$ ps. **(d)** Spatial progression plots the time-variation through a fixed spatial location, which here is rotated coordinates, $x'$, that lines up with one arm of the "X". Progressing in $x'$ causes the line to rotate in space-time. **(e)** The line surface and intersection with $z'$ slices which gives one straight line, instead of two for time slices.

Now, consider measuring the time variation of the field through one of the lines of the "X" oriented at 45 degrees in the $x, z$ plane. In the rotated geometry, equation (1) is seen as

$$\psi = (2x'z' - t + iy)u' \tag{2}$$

Where $x'$ and $z'$ are the rotated coordinates and are defined as $x' = \frac{1}{\sqrt{2}}(x - z)$ and $z' = \frac{1}{\sqrt{2}}(x + z)$. Along one arm of the "X" such as the $x' = 0$ plane, this gives $\psi = (-t + iy)u'$. The null line is at $t = 0$ and as $x'$ increases, the null line tilts in the $z', t$ plane as shown in Figure 1 (d). This means a reconnection between two lines is a single line that follows $2x'z' - t = 0$ which tilts with spatial progression in $x'$. The half-power motion of two lines in time is a result of a single line acquiring tilt linearly with spatial progression.

By symmetry, any line that rotates in time must be two lines reconnecting with progression in an alternative space-time axis. For example, a vortex that develops tilt in $z$ with time,

$$\psi = (2avzt + x + iy)u \tag{3}$$

would also be a reconnection. This is clear from seeing a rotated $\psi$,

$$\psi' = (av(z'^2 - t'^2) + x + iy)u \tag{4}$$

where $z = \frac{1}{\sqrt{2}}(z' - t'), t = \frac{1}{\sqrt{2}}(z' + t')$, where two lines moving in $z'$, reconnect, the move out in $t'$ as $x$ advances.

Now that we have established that two lines reconnecting in time are equivalent to a single rotating line in space, we now show that a cyclically rotating line in time is a reconnection array with spatial progression

$$\psi = (xcos(\omega t) + zsin(\omega t) + iy)u. \tag{5}$$

The null line of the rotating line is $Re(\psi) = 0 \rightarrow \frac{x}{z} = -\tan(\omega t)$, and is plotted with spatial progression in Figures 2 (a) and (b). An array of reconnection occurs in the $x, t$ plane as $z$ progresses. The full null-surface is shown in Figure 2 (c) which reveals a spiral path expected for a rotating line. The $z$ progression lines are plotted on the null surface in Figure 2 (f) which displays the array of reconnections. The rotating line creates the spiral null-surface which has a saddle geometry at each point in time near the spatial center. Thus from a particular spatial progression, there appears to be repeating reconnections stretched across time.

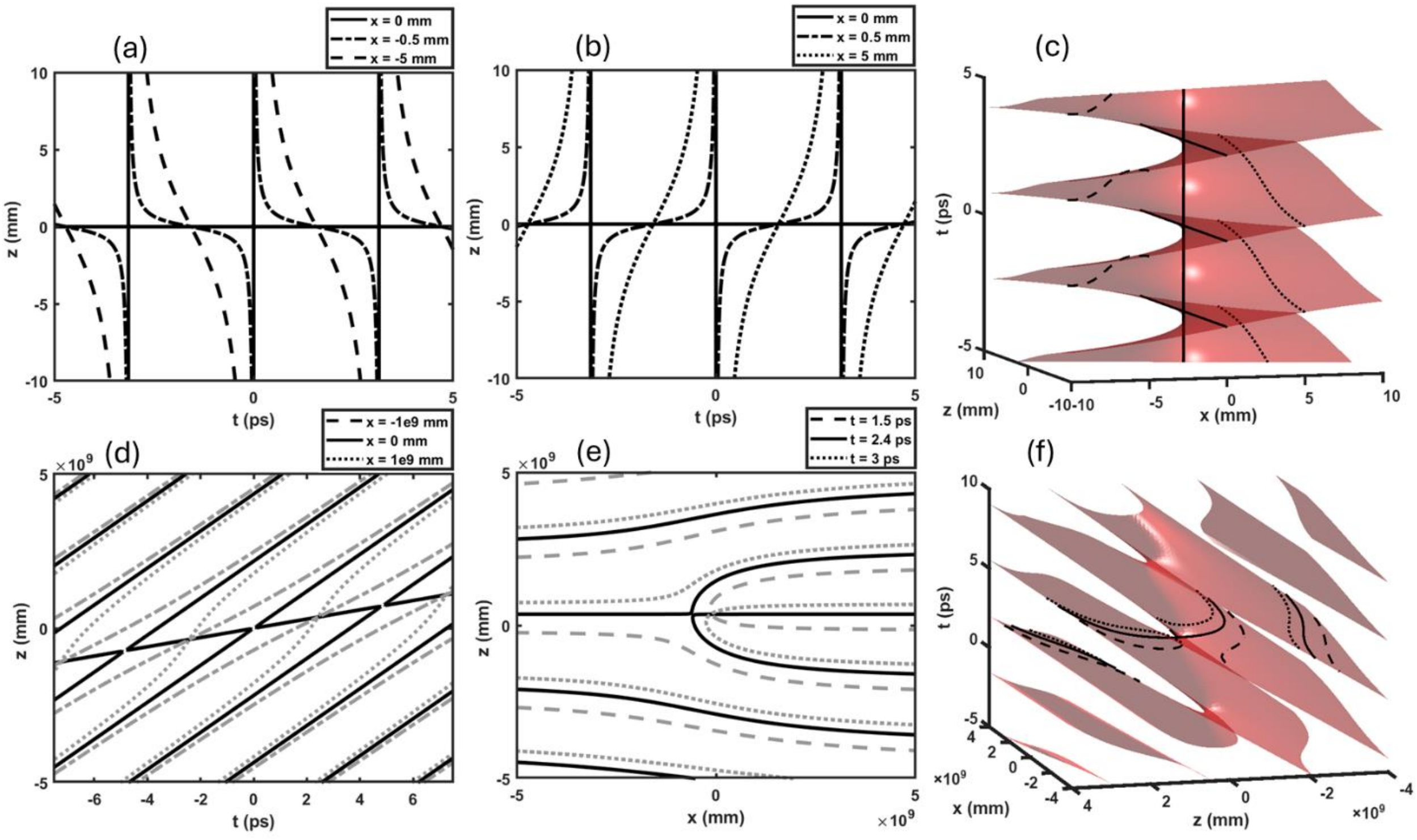

**Figure 2:** A line precessing over time is an array of reconnections over space and, if boosted, cyclic reconnections in time. **(a), (b)** A line rotating in the $x, z$ plane, plotted in $z, t$ for various x values, which shows an array of reconnections. **(c)** The line's trajectory surface and the intersections with $x$ slices. **(d)** The same rotating line seen from a $v = 0.5c$ boosted frame in new z, $t$ coordinates. **(e)** The spatial $x, z$ profile of line plotted for progressing $t$ is now multiple lines that reconnect. **(f)** The line's trajectory surface with corresponding time slices.

From our defined vortex representation, it will be no surprise that a boosted inertial frame, which hyperbolic rotation in space and time, will see the line rotating in time now as a reconnection in time. An inertial frame at velocity $v$ along $z$ would see a null line $\frac{x}{\gamma(z-vt)} = -\tan(\omega\gamma(t - \frac{v}{c^2}z))$, where $\gamma = 1/\sqrt{1 - v^2/c^2}$. The null line is plotted with spatial progression in Figure 2 (d) and with temporal progression Figure 2 (e). While the reconnection array is still present with spatial progression, now from boosted frame, a reconnection appears with temporal progression. The full null-surface is plotted in Figure 2 (f) with $t$ progression lines overlayed.

## 2.2 Magnetic reconnections

### 2.2.1 Methods for magnetic reconnections

Now that we have demonstrated the geometric properties of reconnections and rotating lines, we will show specific examples for physical systems. For the first example, we provide magnetohydrodynamic simulations showing reconnecting magnetic field lines are a single tilting vector potential with a saddle evolution surface. The magnetic-flux contours were generated using a custom two-dimensional Fourier pseudo-spectral solver for incompressible reduced magnetohydrodynamics. The evolved fields were the flux function $A(x, y, t)$ and vorticity $\omega(x, y, t)$, with $J = -\nabla^2 A$ and $\omega = \nabla^2 \phi$, governed by [16,17]

$$\partial_t A = \{\phi, A\} + \nabla \cdot (\eta(\mathbf{r}, t)\, \nabla A) \tag{6}$$

$$\partial_t \omega = \{\phi, \omega\} - \{A, J\} - \nu \nabla^2 \omega \tag{7}$$

where $\{f, g\} = \partial_x f\ \partial_y g - \partial_y f\ \partial_x g$ and viscosity is set to $\nu = 3 \times 10^{-4}$. The resistivity $\eta(\mathbf{r}, t) = \eta_0 + \eta_*(\mathbf{r}, t)$ consists of a uniform base value $\eta_0 = 3 \times 10^{-4}$ plus a time-ramped anomalous enhancement $\eta_*$ with Gaussian spatial distribution and peak $4 \times 10^{-3}$ near the reconnection X-point. The initial condition was a Harris-type current sheet [20]

$$A_0 = B_0 L_{\mathrm{cs}} \ln\left( cosh\left(\frac{y}{L_{cs}}\right)\right) \tag{8}$$

with $B_0 = 4$, $L_{\mathrm{cs}} = 0.5$. Additionally, we included a localized tearing seed $\propto\ \mathrm{sech}^2\left(\frac{y}{L_{\mathrm{cs}}}\right)\cos(k_{x0}x)$ with $k_{x0} = 1/2$, so $k_{x0}L_{\mathrm{cs}} \approx 0.25 < 1$ to ensure the current sheet is tearing unstable and reconnection occurs [21]. The numerical stability is ensured by hyperdiffusion: $\nu_4 \nabla^4 \omega$ with $\nu_4 = 10^{-5}$, drag: $\alpha\omega$ with $\alpha = 1.5 \times 10^{-4}$, and a weakly absorbing boundary layer. Time integration used the Exponential Time Differencing-RK4 [19] and nonlinear terms were evaluated pseudo-spectrally with a 2/3-rule dealiasing mask [22].

### 2.2.2 Magnetic reconnections as a tilting field line

Figures 3 (a) and (b) show contours of the flux function or vector potential $A(x, y, t)$ progressing over time through a reconnection. The full space-time evolution surface of a single contour level is shown in Figure 3 (c). The surface has a saddle shape as in Figure 1 and as a result, slicing this contour in time reveals two field lines reconnecting while slicing in space gives a single field line rotating. For spatial progression, the $x, y$ axes were rotated to $x', y'$ to align with field contours at the reconnection point. Figures 3 (d)-(f) give the spatial progression of $A$ through $y'$ which shows the same reconnection occurring over time which is seen as the magnetic field line tilting as space progresses

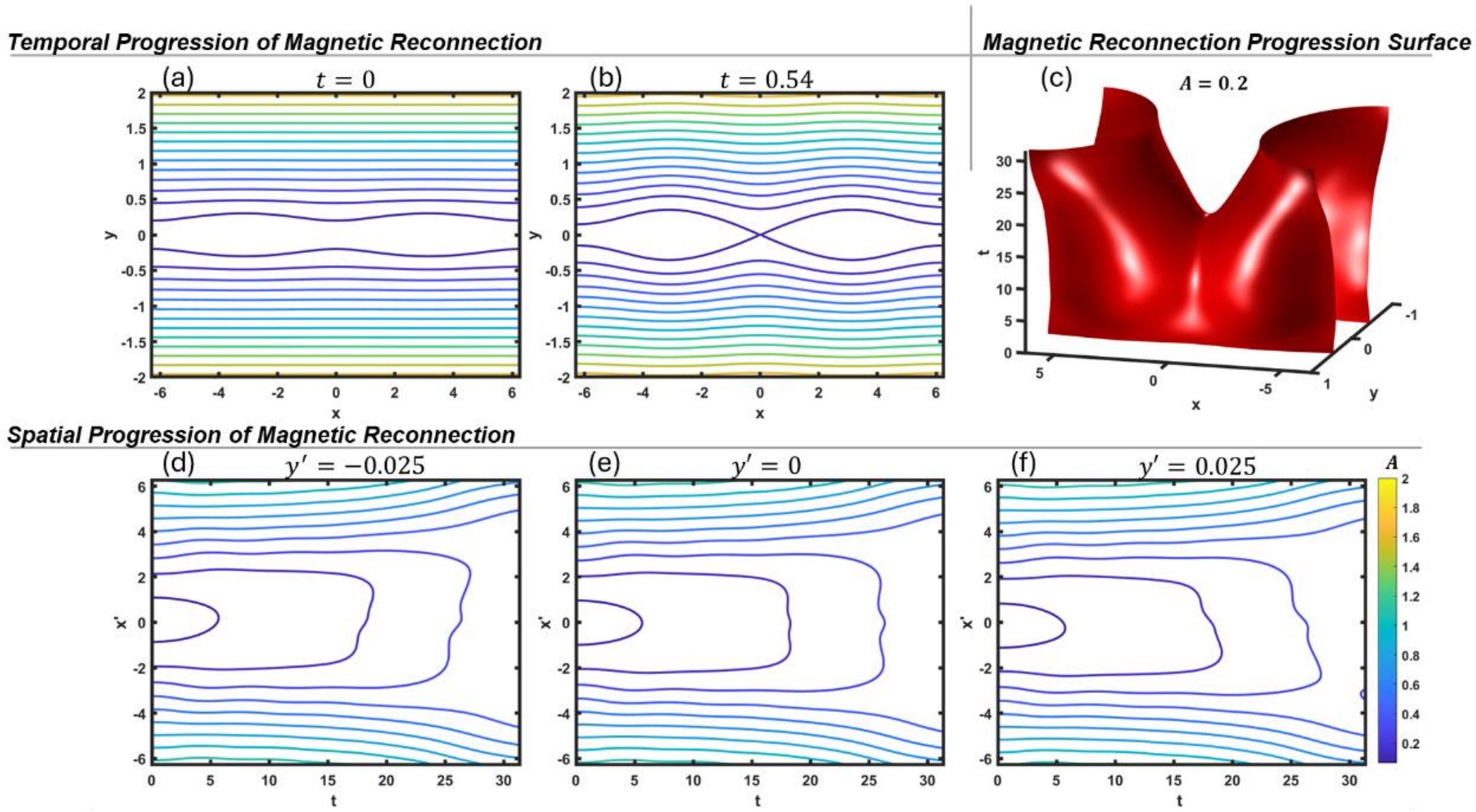

**Figure 3:** A 2D magnetic reconnection simulation. **(a),(b)** Two field lines reconnect with temporal progression. **(c)** The full field line evolution is a surface, which is saddle shaped. **(d)-(f)** A single field line rotates with spatial progression.

A propagating magnetic field line in a plasma wave has a corresponding electric field, which is an extrinsic feature of the reference frame. However, at $y' = 0$ the magnetic reconnection is locally approximately a pure electric field ($\frac{d}{dx'} A_y \approx 0$), as apparent in Figure 3 (e) with the field line being aligned with $x'$. At the center, the extrinsic electric fields seen from the common reference frame add together, while the intrinsic magnetic fields cancel. As $y'$ progresses, the vector potential tilts which changes the electric to magnetic field magnitude ratio.

## 2.3 Electromagnetic free-space vortex precession and reconnections

### 2.3.1 Methods for electromagnetic free-space vortex precession and reconnections

For this section, we first show that both cylindrically focused optical vortices (OVs) and spatiotemporal optical vortices (STOVs) are static reconnections localized wherever the topological charge switches sign. For this analysis, we start with an initial $x$-polarized vector potential $A_0$ and propagate using both angular spectrum method (ASM) without approximation [23]

$$A(\omega, k_x, k_y, z) = A_0 \exp\left( iz\sqrt{\left(\frac{\omega}{c}\right)^2 - k_x^2 - k_y^2} \right) \tag{9}$$

and ASM with a paraxial and narrow bandwidth approximation

$$A(\omega, k_x, k_y, z) \approx A_0 \exp\left( iz\left( \frac{\omega}{c} - \frac{c}{2\omega_o}\left(k_x^2 + k_y^2\right) \right) \right) \tag{10}$$

The approximate propagation equation (eq. 10) allows deriving analytic equations which are adequate for visualizing the vortex fields and is commonly used for spatiotemporal vortex analysis, while the full propagation equation will be shown to be needed to accurately calculate the angular momentum. Additionally, to ensure a divergence free field we apply the standard transverse projector in Fourier space to the $x$-polarized vector potential seed to find the modified $x$ component and remaining field components [24]:

$$A'_i = A_i - k_i(k_j A^j)/k_j k^j \tag{11}$$

Where the $k$ is replaced with $k_0$ for the narrow bandwidth approximation case. Then we calculate electric and magnetic field components

$$E_i = i\omega A_i \tag{12}$$

$$B_i = i\epsilon^{ijk}\partial_j A_k \tag{13}$$

The energy density [25]

$$cp_t = \frac{1}{4} Re\left(\varepsilon_0 E_i E^{*i} + \frac{1}{\mu_0} B_i B^{*i}\right) \tag{14}$$

and linear momentum ($i = \{x, y, z\}$)

$$p_i = \frac{1}{2}\varepsilon_0 Re\left(\epsilon_{ijk} E^j B^{*k}\right) \tag{15}$$

are used to find the 4D angular momentum density with ($i = \{t, x, y, z\}$)

$$L_{ij} = x_i p_j - x_j p_i \tag{16}$$

For example, $L_{tx} = ctp_x - xp_t$ is an angular momentum component associated with boosts while $L_{xy} = xp_y - yp_x$ is an angular momentum component associated with spatial rotations. The total angular momentum, $L^T_{ij}$, is found by integrating over $x, y, z$.

### 2.3.2 Astigmatic focused optical vortices and spatiotemporal optical vortices as static reconnections

Now we will show that astigmatically focused optical vortices and spatiotemporal optical vortices are static reconnections. For weak focused and narrow bandwidth wavepackets, the $x$ polarized electric field visually looks identical to $A$. Additionally, for this analysis, the propagation behavior of $A$ contains all the meaningful vortex reconnection information, while the full divergence free vector field is needed for calculating angular momentum [13]. The input optical vortex field with a cylindrical focusing phase on the $x$ axis is:

$$A = (x + iy)\exp\left(-\frac{x^2}{w^2} - \frac{y^2}{w^2} - \frac{t^2}{\Delta\tau^2} - i\frac{k_0}{2f_x}x^2\right) \tag{17}$$

where $w, \Delta\tau, f_x$, and $k_0$ are the 1/e beam width, pulse width, focal length, and center wavevector. The optical vortex propagated paraxially with narrow bandwidth is

$$\frac{A}{u} = \frac{x}{1 - z\left(\frac{1}{\mathrm{f_x}} - i\frac{2}{k_0 w^2}\right)} + i\frac{y}{1 + iz\,\frac{2}{k_0 w^2}} \tag{18}$$

where $u$ contains the complex envelope and mutual coefficients of the $x, y$ terms of the vortex. Assuming $z \approx f_x$ gives

$$\frac{A}{u'} \approx -\frac{f_x}{z_r}x + y + i\left(\frac{1}{z_r}yz' + x\right) \tag{19}$$

where $z_r = \frac{2f_x^2}{k_0 w^2}$, $z' = z - f_x$ and $u' = u\left(z' + iz\frac{f_x}{k_0 w^2}\right)\left(1 + iz\frac{2}{k_0 w^2}\right)z_r$. This equation reveals a static reconnection that has one of the reconnecting lines tilted in $x, y$. An iso-intensity plot of the full field is given in Figure 4 (a) which also shows the static reconnection for common THz frequency laboratory parameters $w_x = 12.5$ mm, $f_x = 100$ mm, and $f_0 = 0.3\ THz$ with $\omega_0 = 2\pi f_0$. The $x$ coefficient in eq. 19 changes sign as $z$ passes $f_x$ which causes the topological charge of the vortex to change sign through the focus. This topological charge sign change is evidence of the reconnection. The sign change arises from the second vortex that straddles the focus. Figure 4 (d) shows that the normalized total longitudinal angular momentum is unity and constant over time, both for propagating with approximations and without approximations.

On the other hand, a STOV input is

$$A = \left(x + i\frac{w_x}{\Delta\tau}\mathrm{t}\right)\exp\left(-\frac{x^2}{w^2} - \frac{y^2}{w^2} - \frac{t^2}{\Delta\tau^2} - i\frac{k_0}{2f}x^2\right) \tag{20}$$

The approximate field is

$$\frac{A}{u} = \frac{x}{1 - z\left(\frac{1}{\mathrm{f_x}} - i\frac{2}{k_0 \mathrm{w_x^2}}\right)} + i\frac{w_x}{\Delta\tau}\tau \tag{21}$$

where $u$ is the complex envelope and mutual coefficients of $x$ and $\tau = t - z/c$. Intensity cross-sections are plotted in Figure 4 (b) at various times around the focus. The vortex splits up into a Hermite-Gaussian like profile at the focus, then emerges again as a vortex with reverse topological charge after the focus when the $x$ coefficient changes sign. The topological charge sign change is again evidence of a reconnection. Near the focus, $z \approx f_x$ which gives

$$\frac{A}{u'} \approx -\frac{f_x}{z_r}x + \frac{w_x}{\Delta\tau}t' + i\frac{w_x}{\Delta\tau}\frac{1}{z_r}z't' \tag{22}$$

where $t' \approx t - \frac{f_x}{c}$, and common terms between coefficients are grouped into $u' = u\left(z' + iz\frac{f_x}{k_0 w_x^2}\right)z_r$, which don't affect the embedded vortices. For visualization, we use the same $w_x = 12.5$ mm, $f_x = 100$ mm, and $f_0 = 0.3\ THz$ parameters with pulse width $\Delta\tau = 50$ ps. This equation reveals a static reconnection between a $x, z'$ vortex and a $x, t'$ vortex and is visualized in Figure 4 (c). The transverse angular momentum $L_{zx}^T$ and $L_{tx}^T$ are shown in Figures 4 (e) and (f), normalized by $\omega_0/U$ where $U$ is the total energy found by integrating eq. 14 over all space. The approximate analytic field equations give rise to a linear growth in angular momentum, while the field solved numerically without approximations reveal the angular momentum is constant with time. The approximations give rise to a fictitious torque. This torque likely arises from broken $x, z$ and $x, t$ symmetry in the paraxial and narrowband equations, which makes traverse angular momentum not conserved.

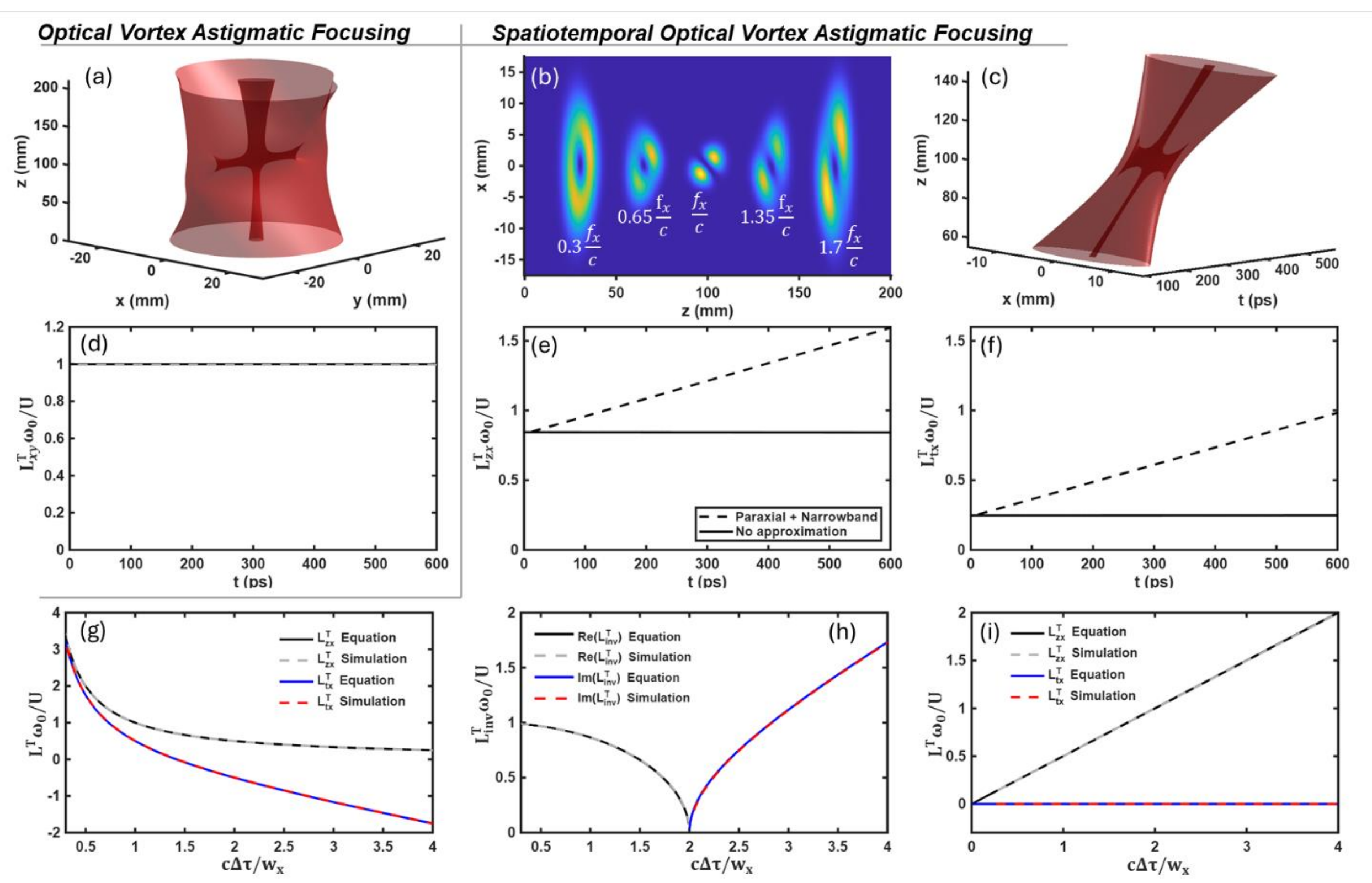

**Figure 4**: Astigmatic propagation of OVs and STOVs are static reconnections. **(a)** Iso-intensity plot of $A$ for the OV and **(d)** the $L_{xy}^{T}$ normalized angular momentum over time for propagating with and without approximations. **(b)** Intensity cross sections of $E_x$ of the STOV (normalized for clarity) and a space-time iso-intensity plot. **(e)** The normalized transverse angular momentum $L_{xz}^{T}$ and **(f)** $L_{xt}^{T}$ over time for propagating with and without approximations. **(g)** Total normalized angular momentum components and **(h)** total boost invariant angular momentum as a function of pulse-to-beam width ratio for the no approximation case. **(i)** Total intrinsic normalized angular momentum components after shifting to the $x, z$ centroids for variable pulse-to-beam width ratio.

Both spatial and spatiotemporal components $L_{zx}^{T}$ and $L_{tx}^{T}$ are non-zero and distinct quantities. As a next step, we look at the property of the field's angular momentum which uses both quantities and does not depend on the inertial frame. $L_{inv}^{2} = |M|^2 - |K|^2$, is an Lorentz invariant [26] where $|M|$ is the norm of the spatial components $(L_{xy}^{T}, L_{zx}^{T}, L_{yz}^{T})$, and $|K|$ the normal of the spatiotemporal components $(L_{tx}^{T}, L_{ty}^{T}, L_{tz}^{T})$. For this case of the STOV, the only non-zero components are $L_{zx}^{T}$ and $L_{tx}^{T}$, so that ${L_{inv}^{T}}^{2} = {L_{zx}^{T}}^{2} - {L_{tx}^{T}}^{2}$. Figure 4 reveals that this is non-zero. However, we can vary the STOV properties to study how this invariant angular momentum value varies and to remove the effect of focusing we set $f_x = \infty$ (Figure 4 (g),(h)). From the data, we found the relationships for the angular momentum values based on the pulse-to-beam width ratio. Figure 4 (g) shows that

$$L_{zx}^{T} \frac{\omega_0}{U} = \frac{w_x}{c\Delta\tau}, \tag{23}$$

$$L_{tx}^{T} \frac{\omega_0}{U} = \frac{w_x}{c\Delta\tau} - \frac{1}{2}\frac{c\Delta\tau}{w_x}, \tag{24}$$

$$L_{inv}^{T} \frac{\omega_0}{U} = \sqrt{1 - \left(\frac{1}{2}\frac{c\Delta\tau}{w_x}\right)^2}, \tag{25}$$

which are plotted in Figure 4 (g) and (h) alongside the simulation calculation. Although $L_{zx}^T \sim L_{tx}^T$ for large beam size and small pulse width, $L_{inv}^T \to 1$ due to the residual $\frac{1}{2}\frac{c\Delta\tau}{w_x}$ in $L_{tx}^T$.

Additionally, we calculated the intrinsic angular momentum by shifting $x$ and $z$ in eq. 16 to the energy centroids $x_c = \frac{\int xcp_t dxdydz}{U}$ and $z_c = \frac{\int zdxdydz}{U}$ at $t = 0$ and plotted in Figure 4 (i). Since $L_{tx}^T(t=0)$ is the $x$ centroid times $U/c$, shifting by the centroid causes the shifted component, $L'_{tx}$, to go to zero. Additionally, with $L'_{zx} = zp_x - xp_z - z_c p_x + x_c p_z$ and $z_c = ct$ and $p_z \sim p_t$ this gives

$$L_{int} \equiv L'_{zx} \cong L_{zx} - L_{tx}. \tag{26}$$

Plugging in eq. 23 and eq. 24 into equation (26) for the total integrals gives the intrinsic angular momentum

$$L_{int}^T \frac{\omega_0}{U} = \frac{1}{2}\frac{c\Delta\tau}{w_x} \tag{27}$$

which is plotted in Figure 4 (i) alongside the simulation calculation for centroid shifted angular momentum.

Eq. 27 matches previous results in the literature for STOVs analyzed with paraxial calculations where the total quantities were calculated and integrated in the local coordinates of the pulse ($i.e. \zeta = z - ct$) [11]. Paraxial calculations may give accurate results before propagation, but any estimate of diverging angular momentum from changing beam size with propagation is fictitious (Figure 4 (e),(f)). Additionally, if we define the angular momentum in unnormalized light-cone coordinates, $\zeta = z - ct, \eta = z + ct$, with momentum component along $\zeta$ as $p_\zeta = \frac{\partial\zeta}{\partial z}p_z + \frac{\partial\zeta}{\partial(ct)}p_t = p_z - p_t$ and $L_{\zeta x} = \zeta p_x - xp_\zeta$, then this gives the local light-cone angular momentum as

$$L_{\zeta x} = L_{zx} - L_{tx}, \tag{28}$$

which matches eq. 26 for intrinsic angular momentum. This shows that the angular momentum in the light cone coordinates can be equivalent to the intrinsic angular momentum found by shifting to the energy centroids. Finally, since $p_z \sim p_t$, $p_\zeta \sim 0$ for paraxial scenarios and $L_{int} \cong L_{x\zeta} \cong \zeta p_x$ which matches the formulation in Ref 9 and gives an alternative perspective why $xp_\zeta$ does not contribute to the angular momentum under certain conditions. However, generally under conditions like high numerical aperture focusing ($p_z \neq p_t$) or wavepacket with $z$ centroids not moving at the speed of light ($z_c \neq ct$), then the intrinsic angular momentum may not equal the light-cone formulation angular momentum.

### 2.3.3 Vortices with spatiotemporal tilt as dynamic reconnections

Now that we have shown that astigmatically propagating OVs and STOVs are static reconnection, we will now demonstrate that an astigmatically focused tilted STOV precesses with spatial progression and is a dynamic reconnection with temporal progression. A tilted STOV input with an $x$ oriented cylindrical lens phase is

$$A = \left(x + i\left(y + \frac{w_x}{\Delta\tau}\tau\right)\right)\exp\left(-\frac{x^2}{w_x^2} - \frac{y^2}{w_y^2} - \frac{t^2}{\Delta\tau^2} - i\frac{k_0}{2f}x^2\right) \tag{29}$$

The analytic approximate field is

$$\frac{A}{u} = \frac{x}{1 - z\left(\frac{1}{\mathrm{f_x}} - i\frac{2}{k_0\mathrm{w_x^2}}\right)} + i\left(\frac{y}{1 + iz\,\frac{2}{k_0\mathrm{w_y^2}}} + \frac{w_x}{\Delta\tau}\tau\right) \tag{30}$$

where $u$ is again the complex envelope and mutual terms. Near the focus, $z \approx f_x$ which gives

$$\frac{A_0}{u'} \approx -\frac{f_x}{z_r}x + y + \frac{w_x}{\Delta\tau}t' + i\left(x + \frac{1}{z_r}\mathrm{yz'} + \frac{w_x}{\Delta\tau}\frac{z_r}{f_x}t'\right) \tag{31}$$

where $z' = z - f_x$, $t' \approx t - f_x/c$, and $u' = u\left(z' + iz\frac{f_x}{k_0\mathrm{w_x^2}}\right)\left(1 - z\left(\frac{1}{\mathrm{f_y}} - i\frac{2}{k_0\mathrm{w_y^2}}\right)\right)z_r$. This equation reveals both a dynamic reconnection with $t$ progression and a tilting vortex with $z'$ progression. Using parameters $w_x = w_y = 12.5$ mm, $\Delta\tau = 100$ ps, $f_0 = 0.3\ THz$, and $f_x = 100$ mm, the field is visualized in Figure 5 (a)-(c) which shows the temporal progression with iso-intensity plot of both $A$ (shown in red) and internal vortices $A/u$ (colored according to phase). While the input vortex was straight, the 3D spatial arrangement of the vortex becomes curved near the focus and a second vortex approaches from the positive $y$ direction (Figure 5(a)). When the energy is localized near the focus, the vortex lines reconnect (Figure 5 (b)) and move apart (Figure 5 (c)).

Spatial progression is shown Figures 5 (d)-(g). Despite the complex 3D vortex having spatial curvature with temporal progression, the tilted spatiotemporal vortex remains straight. The vortex precesses around the $x$ axis and makes a 180° rotation through the focus, as expected from a geometric optics perspective. The rotating trajectory is also seen in Figure 5 (g) which plots the $y, z, t$ iso-intensity of $A/u$ for $x = 5$ mm. At $z = f$, the vortex has no tilt and is a pure spatiotemporal vortex. This zero-tilt vortex is explained from the two colliding vortices which have intrinsic spatial vortices rotating in opposite directions that cancel at this plane, while the extrinsic spatiotemporal vortices rotate in the same direction and add together at this plane.

The total normalized angular momentum terms, both transverse and longitudinal, are plotted in Figure 5 (h). As before, the approximate equations give a fictitious transverse torque while the exact field solved numerically reveals the angular momentum is constant over time.

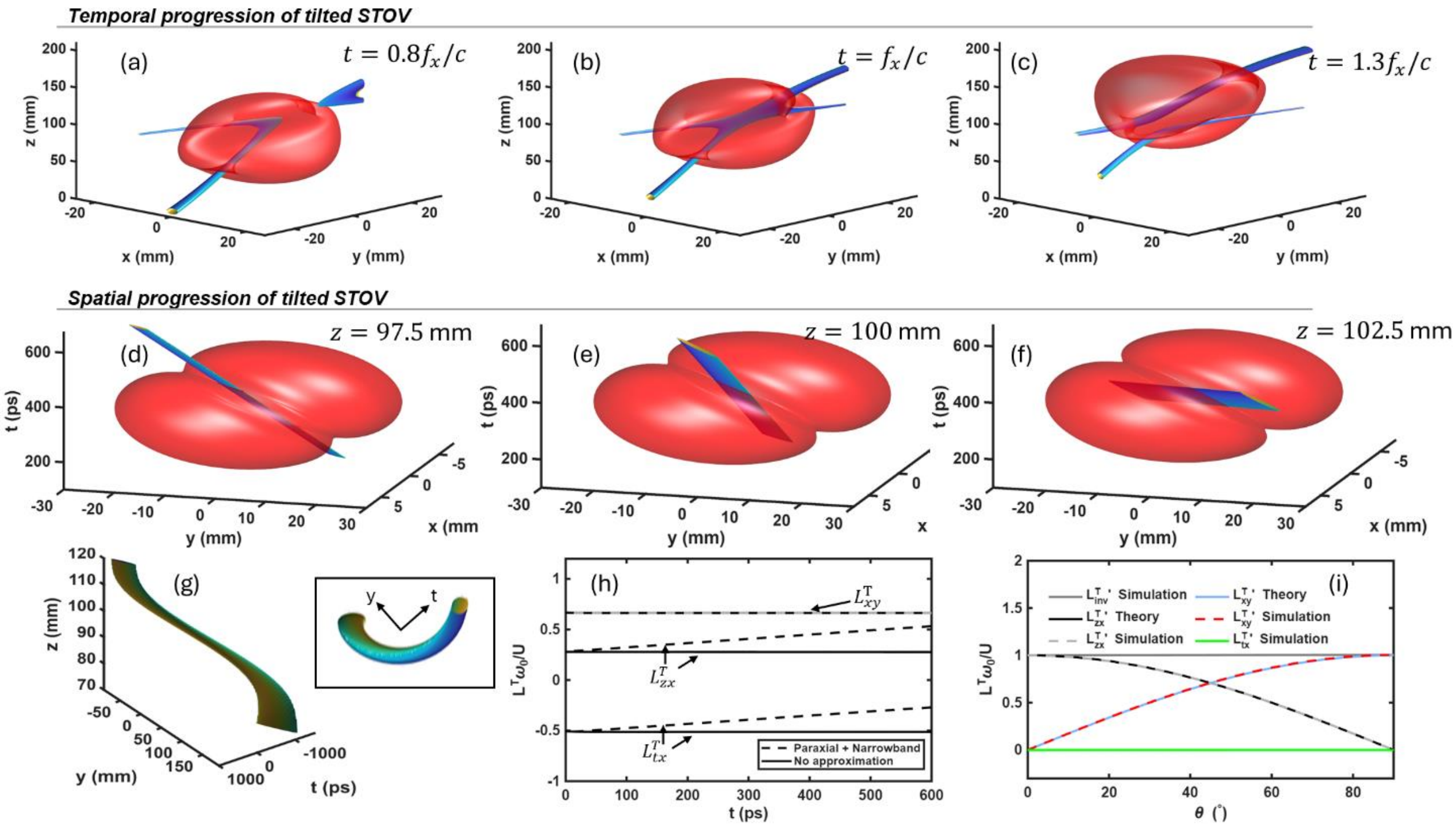

**Figure 5**: Tilted STOVs precess and are dynamic reconnections. (a)-(c) Iso-intensity plots showing temporal progression of a tilted STOV around the focus of $x$ oriented cylindrical lens, revealing a dynamic reconnection. The internal vortex iso-intensity is also shown with phase. (d)-(f) Spatial progression through the focus shows a vortex precessing. (g) An iso-intensity plot of the vortex at $x = 5\ mm$ with an inset top view, showing the rotation of the vortex in $y, t$ as $z$ progresses. (h) Angular momentum components of the field. (i) Intrinsic total angular momentum for variable tilt angle for $c\Delta\tau = 2w_x$.

To study the effect of the tilt angle on the angular momentum components, we can modify the imaginary component of eq. 27 to $ysin(\theta) + \frac{w_x}{\Delta\tau}\tau\cos(\theta)$, where $\theta$ is the tilt angle ranging from 0° (purely transverse) to 90° (purely longitudinal). For this case, we use the parameter $c\Delta\tau = 2w_x$ so that the normalized intrinsic angular momentum is unity at 0° tilt (section 2.3.2). As before, we calculate the intrinsic angular momentum by shifting to the centroids and the results are shown in Figure 5 (i). Additionally, from the data we found that ${L_{xz}^{T}}' = \cos(\theta)$ and ${L_{xy}^{T}}' = \sin(\theta)$ which are plotted alongside the simulation calculations. As ${L_{xt}^{T}}' = 0$ after shifting to the centroids, the intrinsic boost invariant quantity is ${L_{int}^{T}}' = \sqrt{\cos^2(\theta) + \sin^2(\theta)} = 1$ as shown in Figure 5 (i).

In the same manner as section 2.1, we can also characterize the self-similar motion of the vortex lines. Near $x = 0$, the null line follows $\frac{w_x}{\Delta\tau}\frac{z_r^2}{f_x}t' = yz'$. With a 45° rotation in $y, z'$, with $y = \frac{1}{\sqrt{2}}(y_1 - z_1), z' = \frac{1}{\sqrt{2}}(y_z + z_1)$, this becomes $\frac{w_x}{\Delta\tau}\frac{z_r^2}{f_x}t' = \frac{1}{2}(y_1^2 - z_1^2)$. For $t' > 0$, the null line for $z_1 = 0$ follows

$$y_1 = \sqrt{2\frac{w_x}{\Delta\tau}\frac{z_r^2}{f_x}t'} \tag{32}$$

While for $t' < 0$ and $y_1 = 0$, the null line follows

$$z_1 = \sqrt{2\frac{w_x}{\Delta\tau}\frac{z_r^2}{f_x}t'}. \quad (33)$$

Both equations reveal a $t'^{\frac{1}{2}}$ relationship with distance, as expected for a reconnection.

# 3 Conclusion

In conclusion, we have demonstrated that two-vortices reconnecting with time progression are a single spatiotemporal vortex that tilts with space progression. We first presented a toy-model to demonstrate the geometric principle that two lines reconnecting as well as a single line rotating, both possess a saddle shape null-surface oriented in different directions. Two lines reconnecting can be seen as a single line rotating by slicing the null-surface spatially instead of temporally. We then gave an example with a 2D magnetic reconnection simulation, which corroborates the saddle-shape null-surface. Additionally, we showed that a single electromagnetic spatiotemporal vortex with spatiotemporal tilt precesses with spatial progression, and, as a result, can also be seen as a reconnection between two vortices occurring with temporal progression.

# Data availability

Data underlying the results presented in this paper are available from the corresponding authors upon reasonable request.

# Funding

This work was supported by AFOSR under award number FA9550-25-1-0110.

# Competing interests

Authors declare no competing interests.

References

1. Yao, J. and Hussain, F., 2022. Vortex reconnection and turbulence cascade. Annual Review of Fluid Mechanics, 54(1), pp.317-347.
2. Hussain, F. and Duraisamy, K., 2011. Mechanics of viscous vortex reconnection. Physics of Fluids, 23(2).
3. Enciso, A. and Peralta-Salas, D., 2022. Vortex reconnections in classical and quantum fluids. SeMA Journal, 79(1), pp.127-137.
4. Koplik, J. and Levine, H., 1993. Vortex reconnection in superfluid helium. Physical Review Letters, 71(9), p.1375.
5. Hesse, M. and Cassak, P.A., 2020. Magnetic reconnection in the space sciences: Past, present, and future. Journal of Geophysical Research: Space Physics, 125(2), p.e2018JA025935.
6. Le, M.S., Hine, G.A., Goffin, A., Palastro, J.P. and Milchberg, H.M., 2024. Self-focused pulse propagation is mediated by spatiotemporal optical vortices. Physical Review Letters, 133(5), p.053803.
7. Adams, J., Agha, I. and Chong, A., 2024. Spatiotemporal optical vortex reconnections of multi-vortices. Scientific Reports, 14(1), p.5483.

8. Adams, J. and Chong, A., 2025. Spatiotemporal optical vortex reconnections of loop vortices. Nanophotonics, 14(6), pp.729-739.
9. Tripathi, N., Hancock, S.W. and Milchberg, H.M., 2025. Transverse orbital angular momentum: setting the record straight. arXiv preprint arXiv:2503.24375.
10. Hancock, S.W., Tripathi, N., Le, M.S., Goffin, A. and Milchberg, H.M., 2025. Transverse orbital angular momentum of amplitude perturbed fields. Nanophotonics, 14(6), pp.777-784.
11. Porras, M.A., 2024. Clarification of the transverse orbital angular momentum of spatiotemporal optical vortices. Journal of Optics, 26(9), p.095601.
12. Bliokh, K.Y., 2023. Orbital angular momentum of optical, acoustic, and quantum-mechanical spatiotemporal vortex pulses. Physical Review A, 107(3), p.L031501.
13. Adams, J., Park, Y. and Chong, A., 2025. Conservation of transverse orbital angular momentum for spatiotemporal optical vortices. arXiv preprint arXiv:2512.09615.
14. Peskin, M.E., 2018. An Introduction to quantum field theory. CRC press.
15. Barnett, S.M., 2011. On the six components of optical angular momentum. Journal of Optics, 13(6), p.064010.
16. Strauss, H.R., 1976. Nonlinear, three-dimensional magnetohydrodynamics of noncircular tokamaks. The Physics of Fluids, 19(1), pp.134-140.
17. Loureiro, N.F., Schekochihin, A.A. and Cowley, S.C., 2007. Instability of current sheets and formation of plasmoid chains. Physics of Plasmas, 14(10).
18. Cox, S.M. and Matthews, P.C., 2002. Exponential time differencing for stiff systems. Journal of Computational Physics, 176(2), pp.430-455.
19. Kassam, A.K. and Trefethen, L.N., 2005. Fourth-order time-stepping for stiff PDEs. *SIAM Journal on Scientific Computing*, *26*(4), pp.1214-1233.
20. Harris, E.G., 1962. On a plasma sheath separating regions of oppositely directed magnetic field. Il Nuovo Cimento (1955-1965), 23(1), pp.115-121.
21. Furth, H.P., Killeen, J. and Rosenbluth, M.N., 1963. Finite-resistivity instabilities of a sheet pinch. The physics of Fluids, 6(4), pp.459-484.
22. Orszag, S.A., 1971. On the elimination of aliasing in finite-difference schemes by filtering high-wavenumber components. Journal of Atmospheric Sciences, 28(6), pp.1074-1074.
23. K. Matsushima and T. Shimobaba, “Band-limited angular spectrum method for numerical simulation of free-space propagation in far and near fields,” Opt. Express 17, 19662–19673 (2009).
24. Stewart, A.M., 2008. Longitudinal and transverse components of a vector field. arXiv preprint arXiv:0801.0335.
25. J. D. Jackson, Classical Electrodynamics, 3rd ed. (Wiley, 2021).
26. Tung, W.K., 1985. Group theory in physics (Vol. 1). World Scientific.